\documentclass[aps,prb,twocolumn,showpacs,reprint,superscriptaddress]{revtex4}

\usepackage{graphicx}
\usepackage{epstopdf}

\usepackage{times}
\usepackage{amsmath}
\usepackage{amssymb}
\usepackage{amsfonts}
\usepackage{latexsym}

\usepackage{color}

\def\be{\begin{equation}}
\def\ee{\end{equation}}
\def\bea{\begin{eqnarray}}
\def\eea{\end{eqnarray}}

\def\vec{\mathbf}
\def\mc{\mathcal}

\begin{document}
\title{Ferrimagnetism of the magnetoelectric compound Cu$_2$OSeO$_3$ probed by $^{77}$Se NMR}

\author{M. Belesi}\email{mariaeleni.belesi@epfl.ch}\affiliation{Ecole Polytechnique F$\acute{e}d\acute{e}rale$ de Lausanne, ICMP, Station 3, CH-1015 Lausanne-EPFL, Switzerland}
\author{I. Rousochatzakis}\affiliation{Max Planck Institut f$\ddot{u}r$ Physik Komplexer Systeme, N\"othnitzer Str. 38, 01187 Dresden, Germany}
\author{H. C. Wu}\affiliation{CRANN and School of Physics, Trinity College, Dublin 2, Ireland}
\author{H. Berger}\affiliation{Ecole Polytechnique F$\acute{e}d\acute{e}rale$ de Lausanne, ICMP, Station 3, CH-1015 Lausanne-EPFL, Switzerland}
\author{I. V. Shvets}\affiliation{CRANN and School of Physics, Trinity College, Dublin 2, Ireland}
\author{F. Mila}\affiliation{Ecole Polytechnique F$\acute{e}d\acute{e}rale$ de Lausanne, ITP-CTMC, CH-1015 Lausanne, Switzerland}
\author{J. P. Ansermet}\affiliation{Ecole Polytechnique F$\acute{e}d\acute{e}rale$ de Lausanne, ICMP, Station 3, CH-1015 Lausanne-EPFL, Switzerland}

\begin{abstract}
We present a thorough $^{77}$Se NMR study of a single crystal of the magnetoelectric compound Cu$_2$OSeO$_3$.
The temperature dependence of the local electronic moments extracted from the NMR data
is fully consistent with a magnetic phase transition from the high-T paramagnetic phase to a low-T ferrimagnetic state
with 3/4 of the Cu$^{2+}$ ions aligned parallel and 1/4 aligned antiparallel to the applied field of 14.09 T.
The transition to this 3up-1down magnetic state is not accompanied by any splitting of the NMR lines or any abrupt modification in their broadening,
hence there is no observable reduction of the crystalline symmetry from its high-T cubic \textit{P}2$_1$3 space group.
These results are in agreement with high resolution x-ray diffraction and magnetization data on powder samples reported previously by Bos {\it et al.} [Phys. Rev. B, {\bf 78}, 094416 (2008)].
We also develop a mean field theory description of the problem based on a microscopic spin Hamiltonian with one antiferromagnetic ($J_\text{afm}\simeq 68$ K)
and one ferromagnetic ($J_\text{fm}\simeq -50$ K) nearest-neighbor exchange interaction.
\end{abstract}


\maketitle

\section{Introduction}
Multiferroic and magnetoelectric materials are currently at the center of intense research activity.\cite{Eerenstein,Spaldin,Khomskii,Wang}
In magnetoelectric compounds the application of an electric field can induce a finite magnetization and similarly an electric polarization can be induced
by an applied magnetic field.\cite{Landau,Dzyaloshinskii}
Combining electronic and magnetic properties is an exciting playground for fundamental research which may also lead to new multifunctional materials
with potential technological applications. \cite{Khomskii,Gajek,Wood}
Since both time-reversal and spatial-inversion symmetries must be broken in ferroic materials,\cite{Fiebig}
magnetoelectric (ME) effects are allowed only in 58 out of the 122 magnetic point groups.\cite{Fiebig,Schmid}
Along with the weakness of the ME effect, this results in a limited number of compounds displaying ME properties,
such as Cr$_2$O$_3$\cite{Folen}, Gd$_2$CuO$_4$\cite{Wiegelmann}, and BaMnF$_4$\cite{Fox}.
Even though ME materials have been studied extensively in the past decades the recent discovery of the multiferroic compounds TbMnO$_3$\cite{Kimura} and TbMn$_2$O$_5$\cite{Hur}, where ferroelectricity (FE) is driven directly by the spin order, has led to a revival of interest in ME systems.
The magnetic state in these compounds is either a spiral (as in TbMnO$_3$,\cite{Kimura} Ni$_3$V$_2$O$_8$,\cite{Lawes0} and MnWO$_4$\cite{Lautenschl,Taniguchi}),
or a collinear configuration (as in Ca$_3$(CoMn)$_2$O$_6$,\cite{Choi} and FeTe$_2$O$_5$Br\cite{Pregelj}).

Considerable interest has also been drawn to the investigation of ME effects appearing in non-polar systems
such as SeCuO$_3$,\cite{Lawes} TeCuO$_3$,\cite{Lawes} and Cu$_2$OSeO$_3$\cite{Bos}.
The latter, as reported recently by Bos {\it et al.},\cite{Bos}, undergoes a ferrimagnetic phase transition at $T_c\simeq 60$ K which is accompanied by a significant
magnetocapacitance signal and an anomaly of the dielectric constant.  However, high resolution powder x-ray diffraction (XRD) data show that the lattice remains metrically cubic down to 10 K, and this excludes
a ME coupling mechanism that involves a spontaneous lattice strain. This result is further supported by recent infrared,\cite{Miller} and Raman\cite{Gnezdilov} studies. In this respect, Cu$_2$OSeO$_3$ appears to be a unique example of a metrically cubic material that allows for piezoelectric as well as linear magnetoelectric and piezomagnetic coupling.

\begin{figure}[!t]
\centering
\includegraphics[width=0.48\textwidth]{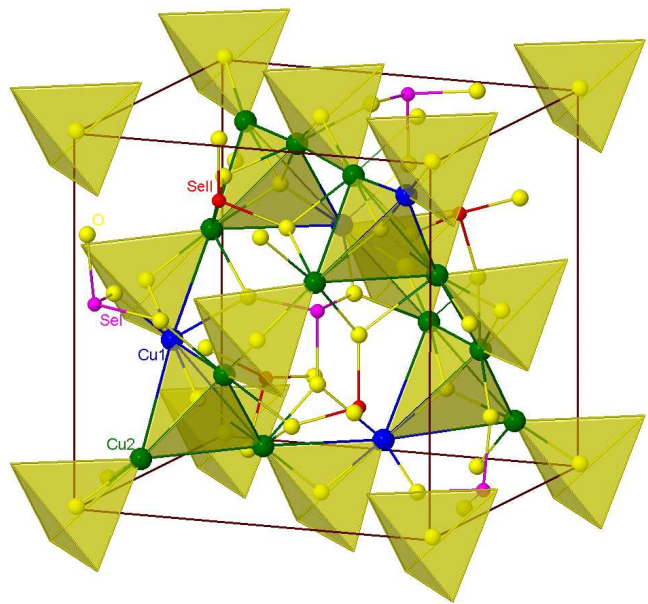}
\caption{(Color online) Crystal structure of Cu$_2$OSeO$_3$. Copper atoms (Cu$_1$ in blue and Cu$_{2}$ in green) form a network of distorted corner sharing tetrahedra. Only the ions inside the unit cell are shown. Oxygen atoms are shown in yellow while Se$_{\textrm{I}}$ and Se$_{\textrm{II}}$ ions are shown in magenta and red respectively.}
\label{CrysStruct}
\end{figure}

Here we present an extensive $^{77}$Se Nuclear Magnetic Resonance (NMR) study which highlights the local magnetic and structural properties of Cu$_2$OSeO$_3$. Our NMR measurements are performed in a single crystal of Cu$_2$OSeO$_3$ as a function of temperature and by varying the direction of the applied magnetic field with respect to the crystalline axes. A detailed analysis of the $^{77}$Se spectra based on space group symmetry considerations leads to the following main conclusions:
The temperature dependence of the local electronic moments extracted from the NMR data is fully consistent with a phase transition between the high-T paramagnetic phase
and a low-T ferrimagnetic configuration with 3/4 of the Cu$^{2+}$ moments (type Cu$_2$) aligned parallel and 1/4 (type Cu$_1$) aligned antiparallel to the applied field.
This 3up-1down state has 1/2 of the saturated magnetization value and is the one proposed by Bos {\it et al}\cite{Bos}.
The transition is not accompanied by any observable change in the broadening of the NMR lines
or any clear splitting, showing that there is no observable symmetry reduction in the crystalline structure
from the high-temperature \textit{P}2$_1$3 space group, in agreement with the reported powder XRD data.\cite{Bos}
In addition, we provide a microscopic description of the problem based on a spin Hamiltonian with two nearest-neighbor exchange couplings:
one antiferromagnetic ($J_\text{afm}\simeq 68$ K) between Cu$_1$ and Cu$_2$ ions, and one ferromagnetic ($J_\text{fm}\simeq -50$ K) between Cu$_2$ ions. The mean field theory predictions of this model are in good qualitative agreement with the behavior of the local moments extracted from NMR and magnetization data.

The article is organized as follows. In Sec. \ref{expdetails} we summarize the synthesis process and give some technical details of our experimental procedure. In Sec. \ref{general} we present our NMR spectra and give a first discussion of some central findings.
The theoretical framework for the explanation of the NMR spectra is given in Sec. \ref{theory} based on symmetry considerations.
The comparison to the NMR data is then given in Sec. \ref{comparison}, where we also extract the relevant transferred hyperfine field parameters as well as the T-dependence of the local moments of both types of Cu$^{2+}$ ions. A comparison to the mean field theory predictions is also made here. Our NMR results for the nuclear spin-lattice $1/T_1$ and spin-spin $1/T_2$ relaxation rates are presented in Sec. \ref{dynamic}. We conclude in Sec. \ref{concl} with a brief discussion of our results.

\section{Synthesis and Experimental details}\label{expdetails}
Single crystals of Cu$_2$OSeO$_3$ were grown by the standard chemical vapour phase method. Mixtures of high purity CuO (Alfa-Aesar, 99.995\%) and SeO$_2$ (Alfa-Aesar, 99.999\%) powder in molar ratio 2:1 were sealed in quartz tubes with electronic grade HCl as the transport gas for the crystal growth. The ampoules were then placed horizontally into a tubular two-zone furnaces and heated very slowly by 50$^\circ$C/h to 600$^\circ$C. The optimum temperatures at the source and deposition zones for the growth of single crystals have been 610$^\circ$C and 500$^\circ$C, respectively. After six weeks, many dark green, almost black  Cu$_2$OSeO$_3$  crystals with a maximum size of 8x6x3 mm were obtained.

The orientation of the crystal axes with respect to the crystal faces were determined by Laue X-ray back-scattering measurements and by single crystal X-ray diffraction. The diffraction data are in agreement with previously published data.\cite{Effenberger,Bos,Larranaga} The crystal structure of Cu$_2$OSeO$_3$ which belongs to the cubic space group \textit{P}2$_1$3 is presented in Fig.~\ref{CrysStruct}. The structure is a three-dimensional array of distorted corner-sharing copper tetrahedra. The unit cell consists of 16 Cu$^{2+}$ ions which belong to two crystallographically different groups denoted here by Cu$_{\text{1}}$ and Cu$_{\text{2}}$, in 4$\alpha$ and 12$\textit{b}$ sites respectively. The oxygen atoms form two different types of distorted CuO$_5$ polyhedra, i.e., trigonal bipyramidal polyhedra for Cu$_{\text{1}}$ sites and square pyramidal polyhedra for Cu$_{\text{2}}$ sites. The CuO$_5$ polyhedra are connected by sharing edges and corners. Similarly, the unit cell contains two crystallographically inequivalent groups of Se$^{4+}$ ions, Se$_{\text{I}}$ and Se$_{\text{II}}$, each one with multiplicity 4. The two types of SeO$_3$ (lone pair) trigonal pyramids share corners with the CuO$_5$ polyhedra.

\begin{figure}[!t]
\centering
\includegraphics[width=0.48\textwidth]{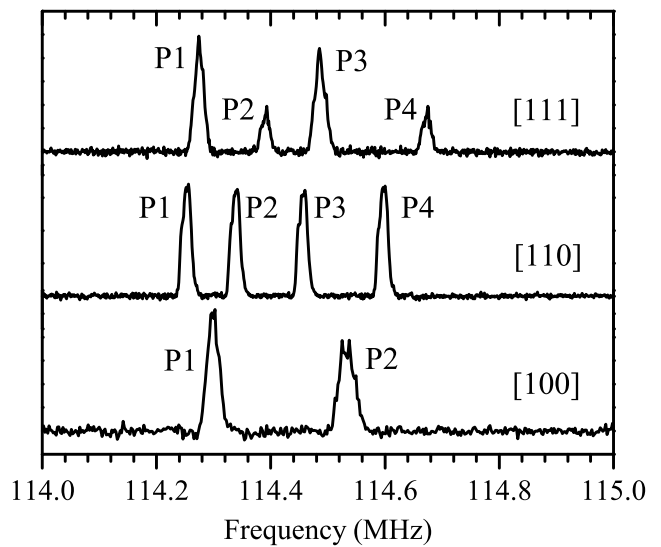}
\caption{$^{77}$Se NMR line shape measurements at 220 K in an external field of 14.09 T applied parallel to the [111], [110] and [100] crystallographic directions.}
\label{Lines220K}
\end{figure}

\begin{table}[!t]
\caption{The experimentally observed paramagnetic shifts $\Delta\nu^{exp}$ (in kHz) at $T=220$ K measured with the external magnetic field $\vec{H}_0$ along the [111], [110] and [100] crystallographic directions.}
\begin{ruledtabular}
\begin{tabular}{c c c c}
 & $\Delta\nu^{exp}_{[111]}$& $\Delta\nu^{exp}_{[110]}$ & $\Delta\nu^{exp}_{[100]}$ \\\hline
$\textrm{P}_{1}$& -121.8$\pm$0.1 & -142.9 $\pm$0.1 & -96.6 $\pm$0.01\\
$\textrm{P}_{2}$& -49.4$\pm$0.2 & -53.9 $\pm$0.1 &   137.5 $\pm$0.04\\
$\textrm{P}_{3}$&  89.9$\pm$0.1 & 60.9  $\pm$0.1 & \\
$\textrm{P}_{4}$&  277.1$\pm$0.2 &  206.8  $\pm$0.1 & \\
\end{tabular}
\end{ruledtabular}
\label{Table220K}
\end{table}

Pulsed NMR experiments were carried out on $^{77}$Se nuclei with nuclear spin $I=1/2$, gyromagnetic ratio $\gamma_{N}/2\pi=8.1179$ MHz/T and natural abundance 7.58$\%$. The NMR experiments were performed on a 125 mg single crystal at an Oxford 14.09 T magnet equipped with a variable temperature insert utilizing a home built spectrometer. The echo signal was produced by a standard Hahn echo pulse sequence with a typical  $\pi/2$  pulse length of 6 $\mu s$. The separation between the two echo-generating pulses was 20$-$60 $\mu s$ depending on the experimental conditions. The $^{77}$Se NMR spectra were measured by Fourier transform (FT) of the half spin-echo signal whenever the whole line could be irradiated with one radiofrequency pulse. For the broad lines, the spectrum was obtained by plotting the area of the echo as a function of the irradiation frequency (frequency sweep). The nuclear spin-lattice $T_1$ and  spin-spin $T_2$ relaxation times were measured using the standard spin echo pulse sequence combined with the saturation recovery method for $T_1$ measurements. For the analysis of the NMR data we have measured the magnetization of a 28.86 mg single crystal of Cu$_2$OSeO$_3$ at 14 T using a Quantum Design PPMS (Physical Properties Measurement System) apparatus located in the Trinity College in Dublin, Ireland.

\section{$^{77}$Se Nuclear Magnetic Resonance spectra}
\subsection{General results}\label{general}
The nuclear $^{77}$Se spins provide a powerful local probe of the behavior of the electronic Cu$^{2+}$ moments
through the transferred hyperfine and the magnetic dipolar mechanisms.
Here we have performed detailed $^{77}$Se NMR lineshape measurements at 14.09 T
with the magnetic field applied parallel to the crystallographic directions [111], [110] and [100].
Figure \ref{Lines220K} shows the $^{77}$Se NMR spectrum at $T=220$ K.
There appear four distinct resonance lines when the field is applied along [111] and [110],
while two lines are observed along the [100] direction. The specific values of the shifts of the lines $\Delta\nu=\nu-\nu_{L}$ from the bare Larmor frequency $\nu_{L}$
are provided in Table~\ref{Table220K}.
The integrated areas under these lines are found with the relative ratios 3:1:3:1 for [111], 2:2:2:2 for [110], and 4:4 for [100].
Since the area under a given line is proportional to the number of the nuclei that resonate in the corresponding frequency window, we conclude that there are four
magnetically inequivalent groups of $^{77}$Se sites in the [111] and [110] directions and only two groups for [100].
It is clear that the latter correspond to the two crystallographically inequivalent groups of selenium sites Se$_\textrm{I}$ and Se$_\textrm{II}$ which have the same multiplicity (four per unit cell). On the other hand, both Se$_{\textrm{I}}$ and Se$_{\textrm{II}}$ groups split into two magnetically inequivalent subgroups when the field is along [111] and [110] with multiplicities 3:1 and 2:2 respectively. Below in Sec. \ref{theory}, we shall be able to identify these specific subgroups of Se sites (and even retrieve the values of the corresponding elements of the transferred hyperfine tensor) by taking into account the local symmetry around each Se site under the conditions that (1) we are in the 3up-1down state and (2) that the crystalline structure belongs to \textit{P}2$_1$3 space group. The second condition is an important issue since any symmetry reduction of the crystal from \textit{P}2$_1$3 to either of the two possible crystallographic subgroups \textit{R}3 or \textit{P}2$_1$2$_1$2$_1$,\cite{IntTablesCryst} will result in partially or fully splitting of the above multiple lines.

\begin{figure}[!t]
\centering
\includegraphics[width=0.49\textwidth]{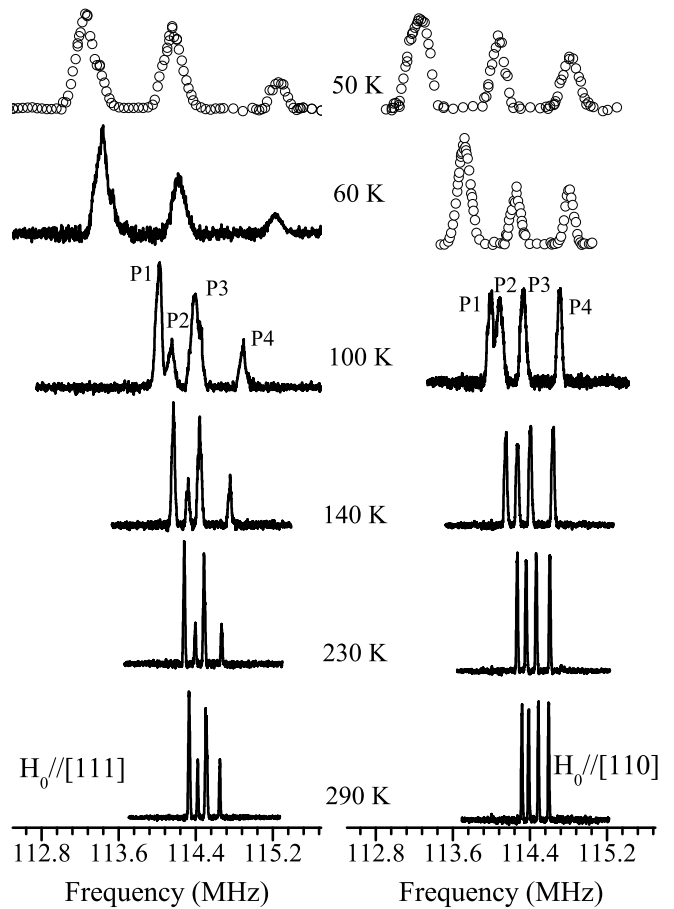}
\caption{$^{77}$Se NMR line shape measurements for selected temperatures measured at external field of 14.09 T applied parallel to the [111] and [110] crystallographic directions.}
\label{LinesVsT}
\end{figure}

\begin{figure}[!t]
\centering
\includegraphics[width=0.48\textwidth]{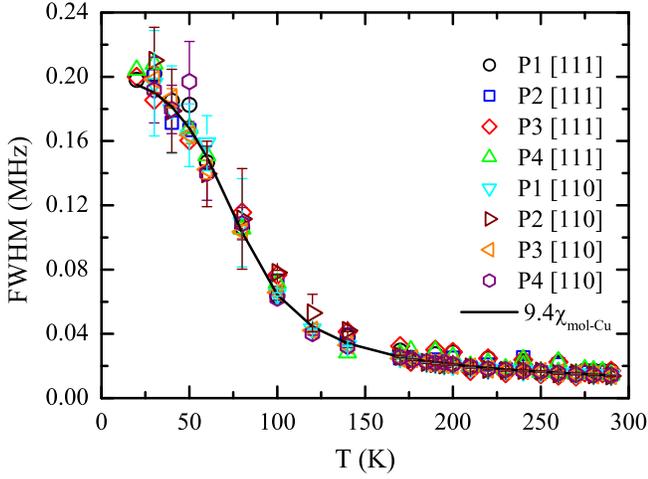}
\caption{(Color online) Temperature dependence of the $^{77}$Se NMR full width at half maximum (FWHM) for each of the four spectral lines measured at 14.09 T along the [111] and [110] crystallographic directions. The FWHM for lines P1 and P3 along [111] are multiplied by a factor 0.9 and 0.7 respectively in order to match the other curves. The temperature dependence of the magnetic susceptibility per copper mol (multiplied by a factor of 9.4) as measured with the magnetometer at 14 T is shown with a solid line.}
\label{FWHM}
\end{figure}

Let us now examine what happens as we cool down the system. We have studied the temperature dependence of the NMR spectra in the range 20-290 K with the magnetic field applied along [111] and [110]. Some representative NMR spectra are shown in Fig.~\ref{LinesVsT}.
We first note that the lines P1 and P2 get gradually closer to each other below 100 K and merge into a single peak at lower temperatures. We also find a gradual increase in the line broadenings or Full Width at Half Maximum (FWHM). However, as we show in Fig. \ref{FWHM}, all the NMR line broadenings (shown with symbols in Fig. \ref{FWHM}) follow quite closely the corresponding gradual increase of the magnetization as measured by PPMS (solid line). This being the typical behavior of inhomogeneous broadening, together with the fact that we find no observable splitting of any of the multiple lines, we are lead to conclude that there is no clear sign of any symmetry reduction of the crystalline structure as the system enters the ferrimagnetic state.

\begin{figure}[!t]
\centering
\includegraphics[width=0.45\textwidth]{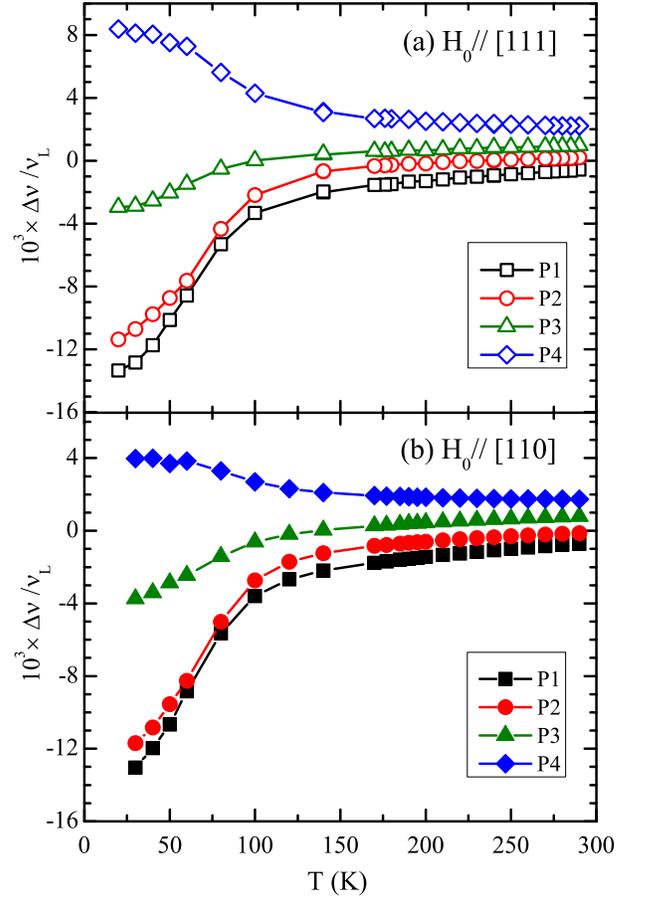}
\caption{(Color online) Temperature dependence of the relative NMR shifts $\Delta\nu/\nu_{L}$ at H$_0=14.09$ T and for each of the four spectral lines observed along the (a) [111] and (b) [110] crystallographic directions. }
\label{RelativeShift}
\end{figure}

Another important issue is the behavior of the line shifts as we cool down the system, since this is governed by the corresponding ordering behavior of the electronic Cu$^{2+}$ moments. Figure \ref{RelativeShift} shows the T-dependence of the relative shifts $\Delta\nu/\nu_{L}$ measured at H$_0=14.09$ T for fields applied along the directions [111] and [110]. From these data, as we show below in Sec. \ref{theory}, we shall be able to extract the T-dependence of the local moments of both Cu$_{\text{1}}$ and Cu$_{\text{2}}$ ions and thus confirm the ferrimagnetic nature of the low-T phase of this compound.

\subsection{Theoretical analysis}\label{theory}
In order to understand and analyze the above NMR data we first note that the most relevant terms in the nuclear spin-1/2 Hamiltonian for the $k$-th Se site are
(i) the Zeeman coupling with the external field $\vec{H}_0$, (ii) the transferred-hyperfine interactions with the relevant neighboring Cu spins
and (iii) the dipolar interactions with all the electronic Cu spins, namely
\be\label{Ham}
\mc{H}_k=-\gamma_{N} \hbar \vec{I}_k \cdot \left( \left(1+\vec{K}^c_k\right)\cdot\vec{H}_0 +\vec{H}_k^{\textrm{tr}}+\vec{H}_k^{\textrm{dip}} \right)~,
\ee
where $\vec{K}^c_k$ denotes the chemical shift tensor, while $\vec{H}_k^{\textrm{tr}}$ and $\vec{H}_k^{\textrm{dip}}$ are the local effective fields resulting from the transferred-hyperfine and the dipolar interactions respectively.

\begin{figure}[!t]
\centering
\includegraphics[angle=0,width=0.7\linewidth]{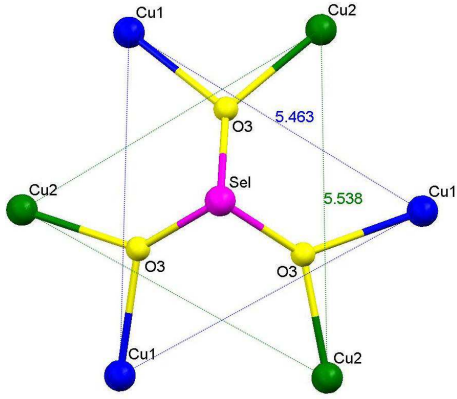}
\caption{(Color online) The local environment of each Se$_\textrm{I}$ nuclear site. There is a 3-fold symmetry axis through the Se site which is vertical to the planes of the Cu$_\text{1}$ and Cu$_\text{2}$ ions. Note that the plane of the Cu$_\text{2}$ ions lies slightly above the plane of the Cu$_\text{1}$ ions. Shown distances are in units of $\AA$.}
\label{StructSe1}
\end{figure}

\textit{Transferred-hyperfine contribution}.---
Let us discuss the transferred-hyperfine coupling first. The transferred-hyperfine interaction is local, i.e. it involves a $^{77}$Se nuclear spin and its six nearest neighbor Cu ions, and is mediated through the Cu-O-Se bonding. The local transferred-hyperfine field experienced by the $k$-th nuclear spin can be written as
\be
\vec{H}^{\textrm{tr}}_k=q \sum_{j} \vec{A}_k^j \cdot \vec{M}_j~,
\ee
where $q \equiv 1/(\gamma_N\hbar g \mu_B)$, $\vec{M}_j=-g\mu_B\langle \vec{S}_j \rangle$ denote the local moments of the six neighbor Cu ions,
$\mu_B$ is Bohr's magneton, and $g$ is the spectroscopic factor with $g\simeq 2.11$.\cite{Larranaga}
To go beyond this general description we need to exploit the local symmetry properties of the electronic environment of each Se site. As we mentioned in the introduction, each unit cell contains four crystallographically equivalent sites of Se$_\textrm{I}$ ions and similarly four equivalent sites of Se$_\textrm{II}$ ions. We denote these by Se$_\textrm{I}^n$ and Se$_{\textrm{II}}^n$ respectively, where $n=1,\ldots,4$. We start our analysis with the first type of selenium ions. As can be seen in Fig.~\ref{StructSe1}, each Se$_{\textrm{I}}^n$ ion sits on a high symmetry crystal site. Each oxygen of the trigonal pyramid SeO$_3$ is connected to two different types of copper ions, denoted as Cu$_1$ and Cu$_2$. The three Se$_{\textrm{I}}$-O-Cu$_1$-Cu$_2$ bonding groups are equivalent and are mapped to one another by a rotation of $120^{\circ}$ around a 3-fold symmetry axis. This $\mathcal{C}_3$ axis passes through the selenium site and is vertical to the plane of the three Cu$_2$ ions. This plane is parallel and slightly above the plane formed by the three Cu$_1$ ions, as well as to the plane formed by the three oxygen atoms (see Fig.~\ref{StructSe1}).
Thus the transferred hyperfine interaction at any given Se$_\textrm{I}^n$ site can be described by two hyperfine tensors, $\vec A^{\textrm{Cu}_1}_{\textrm{Se}^n_{\textrm{I}}}$ and $\vec A^{\textrm{Cu}_{2}}_{\textrm{Se}^n_\textrm{I}}$. The former (latter) represents the sum of the three hyperfine tensors between the $n$-th Se$_\textrm{I}$ and the three Cu$_1$  (Cu$_2$) ions. Now, one of the two 3-fold rotations reads $(x,y,z)\mapsto (z,x,y)$, and this transforms the elements of any second rank tensor as
\be
\begin{pmatrix}
xx & xy & xz\\
yx & yy & yz\\
zx & zy & zz\\
\end{pmatrix}\mapsto
\begin{pmatrix}
zz & zx & zy\\
xz & xx & xy\\
yz & yx & yy \\
\end{pmatrix}.
\ee
Since this is a symmetry operation we must have $xx=yy=zz$ and $xy=xz=yz$.
These conditions describe an axially symmetric tensor with the third principal eigenvector along the [111] direction.
Thus the hyperfine tensors $\vec A^{\textrm Cu_1}_{\textrm Se^n_{1}}$ and $\vec A^{\textrm Cu_{2}}_{\textrm Se^n_\textrm{I}}$
are also axially symmetric and their principal axes system can be assigned easily.
One is the 3-fold axis ($\vec{e}_{3}$) while for the remaining two we may take any pair of mutually perpendicular axes
(denoted by $\vec{e}_{1}$ and $\vec{e}_{2}$) on the plane formed by the three Cu$_2$ (or Cu$_1$) neighboring ions.
Thus we may write
\be\label{Se1Cu1Cu2matrices}
q\mathbf{A^{\textrm Cu_{1}}_{\textrm Se^n_\textrm{I}}} =
\begin{pmatrix}
\alpha_{\textrm{tr}} & 0& 0\\
0 & \alpha_{\textrm{tr}}& 0\\
0& 0& \beta_{\textrm{tr}}\\
\end{pmatrix},
q\mathbf{A^{\textrm Cu_{2}}_{\textrm Se^n_\textrm{I}}} =
\begin{pmatrix}
\alpha'_{\textrm{tr}} & 0& 0\\
0 & \alpha'_{\textrm{tr}} & 0\\
0& 0& \beta'_{\textrm{tr}} \\
\end{pmatrix}~,
\ee
where we emphasize that each tensor is written in its own principal axes frame.
In what follows the above diagonal elements are treated as fitting parameters.

\begin{figure}[t]
\centering
\includegraphics[width=0.95\linewidth]{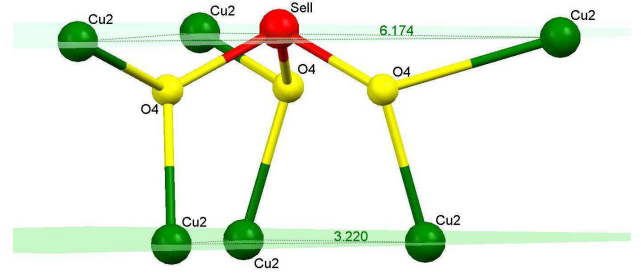}
\caption{(Color online) The local environment of Se$_\textrm{II}$ type ions. As in the case of Se$_{\text{I}}$ ions,  there is a 3-fold symmetry axis through the Se site which is vertical to the planes of Cu$_\text{2}$ ions. Shown distances are in units of $\AA$.}
\label{SrtuctSe2}
\end{figure}

We now turn to the local symmetry of the Se$_\textrm{II}^n$ ions.
These are bonded through oxygen sites to six Cu$_2$ ions (Fig.~\ref{SrtuctSe2}) which belong to two parallel planes.
As in the case of the Se$_\textrm{I}^n$ sites, a 3-fold rotation axis is again present and this passes through the selenium site of the trigonal pyramid SeO$_3$
and is vertical to the set of planes defined by the Cu$_2$ ions. The hyperfine interactions at each Se$_2^n$ site can thus be described with one hyperfine tensor, $\vec A^{\textrm Cu_2}_{\textrm Se^n_\textrm{II}}$, which stands for the sum of the hyperfine tensors of all six neighboring Cu$_2$ ions.
Due to the 3-fold axis this tensor is axially symmetric and, as explained for the Se$_\textrm{I}$ case, the principal axes are the 3-fold axis
and the two mutually perpendicular axes (their plane is parallel to the two Cu planes).
Written in its own principal axis system, the hyperfine tensor $\vec A^{\textrm Cu_2}_{\textrm Se^n_\textrm{II}}$ can thus be written as \be\label{Se2Cu2matrice}
q\mathbf{A^{\textrm Cu_{2}}_{\textrm{Se}_\textrm{II}^n}} =
\begin{pmatrix}
\delta'_{\textrm{tr}} & 0& 0\\
0 & \delta'_{\textrm{tr}} & 0\\
0& 0& \vartheta'_{\textrm{tr}} \\
\end{pmatrix}~.
\ee
Summarizing, the hyperfine tensors for each one of the four Se$^n_\textrm{I}$ sites (respectively Se$^n_\textrm{II}$ sites) of the unit cell have the form of Eqs.~\ref{Se1Cu1Cu2matrices} (resp. \ref{Se2Cu2matrice}) as long as they are written in their own local principal axis coordinate frame.
In Table~\ref{TabSeCoordBloc} we provide the coordinates of the unit vector $\vec{e}_{3}$ in the (x,y,z) frame for each of the eight selenium sites in the unit cell.
The coordinate frames of the hyperfine tensors for the four Se$^n_\textrm{I}$ and the four Se$^n_\textrm{II}$ sites are related with proper rotation of their principal axes $[\vec{e}_{1},\vec{e}_{2},\vec{e}_{3}]$.

\textit{Dipolar contribution}.--- In order to provide a more accurate quantitative description of the measured NMR spectra we must also consider the dipolar coupling.
In contrast to the transferred-hyperfine interactions, the dipolar coupling is long-ranged and since we are dealing with a ferrimagnet
we must also consider the effect of the demagnetization field. To this end we follow the Lorentz method\cite{White} which consists
of splitting the summation over the dipolar contributions into two parts. In the first we perform a discrete summation over all individual
dipolar contributions from all Cu ions enclosed in a sphere of radius $R$ much larger than the lattice spacing and much smaller than the size of the sample.
The summation outside this Lorentz sphere gives rise to the demagnetization field $\vec{H}^{\text{dem}}$ which can be evaluated using a continuous integration.
This gives the well known expression\cite{White} $\vec{H}^{\text{dem}}=4\pi (\frac{1}{3}-\vec{N}_d)\cdot \widetilde{\vec{M}}$, where the demagnetizing tensor $\vec{N}_d$ is specific to the shape of the sample and $\widetilde{\vec{M}}= \left( 4 \vec{M}_1+12 \vec{M}_2 \right)/a^3=4 \vec{M} /a^3$ is the total moment divided by the volume of the sample.
The demagnetizing tensor is not known but we shall be able to adjust it so that we obtain a reasonable T-dependence
of the extracted moments at high temperatures (cf. below).

Following the above, the total dipolar field at the $k$-th Se nuclear site can be written as
\be\label{Bdip}
\vec{H}^{\textrm{dip}}_k=q\sum_{r_{jk} \le R} \vec{D}_k^j\cdot \vec{M}_j + \vec{H}^{\text{dem}}~,
\ee
where as usually the cartesian components ($\alpha, \beta =x,y,z$) of the dipolar tensors are given by
\be\label{DipTensor}
q\left( \vec{D}^j_k \right)_{\alpha\beta} =-\frac{\delta^{\alpha\beta}}{r_{jk}^3}+3\frac{\vec{r}_{jk}^{\alpha} \vec{r}_{jk}^{\beta}}{r_{jk}^5}~,
\ee
where $\vec{r}_{jk}$ stands for the displacement vector from the $k$-th nuclear site to the $j$-th electronic spin.
Summing over the two different types of Cu$^{2+}$ ions in Eq. (\ref{Bdip}) separately we get
\be
\vec{H}^{\textrm{dip}}_k=  q\left( \vec{D}_k^{\textrm{Cu}_1}\cdot \vec{M}_1+ \vec{D}_k^{\textrm{Cu}_2}\cdot \vec{M}_2 \right) + \vec{H}^{\text{dem}}~.
\ee
Using the known values\cite{Effenberger,Bos,Larranaga} of the positions of the Cu$^{2+}$ moments one may evaluate the summations over the Lorentz sphere with very good accuracy (we find that the sums converge already for $R\sim 5 a$). Now, the symmetry of the crystal necessitates that the resulting dipolar tensors have the same principal axes as the corresponding transferred-hyperfine tensors. Indeed written in the corresponding local coordinate frame we find
{\setlength\arraycolsep{0.5pt}
\bea
q\mathbf{D^{\textrm Cu_{1}}_{\textrm{Se}_\textrm{I}^n}}&=&
\begin{pmatrix}
\alpha_{\textrm{dip}} & 0& 0\\
0 & \alpha_{\textrm{dip}} & 0\\
0& 0& \beta_{\textrm{dip}}\\
\end{pmatrix},
q\mathbf{D^{\textrm Cu_{2}}_{\textrm{Se}_\textrm{I}^n}}=
\begin{pmatrix}
\alpha'_{\textrm{dip}} & 0& 0\\
0 & \alpha'_{\textrm{dip}} & 0\\
0& 0& \beta'_{\textrm{dip}}\\
\end{pmatrix},\nonumber\\
q\mathbf{D^{\textrm Cu_{1}}_{\textrm{Se}_\textrm{II}^n}}&=&
\begin{pmatrix}
\delta_{\textrm{dip}} & 0& 0\\
0 & \delta_{\textrm{dip}} & 0\\
0& 0& \vartheta_{\textrm{dip}}\\
\end{pmatrix},
q\mathbf{D^{\textrm Cu_{2}}_{\textrm{Se}_\textrm{II}^n}}=
\begin{pmatrix}
\delta'_{\textrm{dip}} & 0& 0\\
0 & \delta'_{\textrm{dip}} & 0\\
0& 0& \vartheta'_{\textrm{dip}}\\
\end{pmatrix}~,
\eea
}
where, in units of $10^{22} \textrm{cm}^{-3}$:
$\alpha_{\textrm{dip}}=1.68$, $\beta_{\textrm{dip}}=-3.37$, $\alpha'_{\textrm{dip}}=1.18$, $\beta'_{\textrm{dip}}=-2.36$,
$\delta_{\textrm{dip}}=-0.58$,  $\vartheta_{\textrm{dip}}=1.17$, $\delta'_{\textrm{dip}}=-1.3$, and $\vartheta'_{\textrm{dip}}=2.61$.

\begin{table}[!t]
\caption{The second column gives the unit vector $\vec{e}_{3}$ in the fixed (x,y,z) frame for each of the eight Se sites per unit cell (first column).
The following columns give the theory predictions for the sum of the dipolar (excluding the demagnetization field) and the transferred hyperfine contributions to the relative NMR shifts, when the field is along the [111], [110], and [100] crystallographic directions. The dimensionless parameters $\widetilde{\alpha}, \widetilde{\beta}, \widetilde{\delta},\widetilde{\vartheta}$ are defined in Eqs. (\ref{alpha})-(\ref{theta}).}\label{TabSeCoordBloc}
\begin{ruledtabular}
\begin{tabular}{c c c c c}
Se                         & $\vec{e}_{3}$ & [111] &  [110] & [100]\\
\hline
Se$^1_\textrm{I}$ & (1,1,1)/$\sqrt{3}$   & $\widetilde{\beta}$   & $(\widetilde{\alpha}+2\widetilde{\beta})/3$  & $(2\widetilde{\alpha}+\widetilde{\beta})/3$ \\
Se$^2_\textrm{I}$ & (1,1,-1)/$\sqrt{3}$ & $(8\widetilde{\alpha}+\widetilde{\beta})/9$ &$(\widetilde{\alpha}+2\widetilde{\beta})/3$ & $(2\widetilde{\alpha}+\widetilde{\beta})/3$\\
Se$^3_\textrm{I}$ & (-1,1,1)/$\sqrt{3}$ & $(8\widetilde{\alpha}+\widetilde{\beta})/9$  &$\widetilde{\alpha}$ 			      & $(2\widetilde{\alpha}+\widetilde{\beta})/3$ \\
Se$^4_\textrm{I}$ & (1,-1,1)/$\sqrt{3}$ & $(8\widetilde{\alpha}+\widetilde{\beta})/9$  &$\widetilde{\alpha}$ 			      & $(2\widetilde{\alpha}+\widetilde{\beta})/3$\\
\hline
Se$^1_\textrm{II}$ & (1,1,1)/$\sqrt{3}$  & $\widetilde{\vartheta}$ &$(\widetilde{\delta}+2\widetilde{\vartheta})/3$ & $(2\widetilde{\delta}+\widetilde{\vartheta})/3$   \\
Se$^2_\textrm{II}$ & (1,1,-1)/$\sqrt{3}$ & $(8\widetilde{\delta}+\widetilde{\vartheta})/9$  &$(\widetilde{\delta}+2\widetilde{\vartheta})/3$  & $(2\widetilde{\delta}+\widetilde{\vartheta})/3$\\
Se$^3_\textrm{II}$ & (-1,1,1)/$\sqrt{3}$ & $(8\widetilde{\delta}+\widetilde{\vartheta})/9$  &$\widetilde{\delta}$    & $(2\widetilde{\delta}+\widetilde{\vartheta})/3$ \\
Se$^4_\textrm{II}$ & (1,-1,1)/$\sqrt{3}$ & $(8\widetilde{\delta}+\widetilde{\vartheta})/9$  &$\widetilde{\delta}$    & $(2\widetilde{\delta}+\widetilde{\vartheta})/3$ \\
\end{tabular}
\end{ruledtabular}
\end{table}

\textit{Predictions for the relative shifts}.---
Including both the transferred-hyperfine and the dipolar contributions it is straightforward to deduce the relative shifts for each Se site and for each direction of the applied field considered in our experiments. The resulting expressions -- without the contribution for the demagnetization field and the chemical shift -- are provided in Table \ref{TabSeCoordBloc}  in terms of the dimensionless parameters
\bea
\label{alpha} \widetilde{\alpha}&=& (\alpha_{\textrm{tr}}+\alpha_{\textrm{dip}}) \chi_1 + (\alpha'_{\textrm{tr}}+\alpha'_{\textrm{dip}})  \chi_2\\
\label{beta} \widetilde{\beta}&=& (\beta_{\textrm{tr}}+\beta_{\textrm{dip}}) \chi_1 + (\beta'_{\textrm{tr}}+\beta'_{\textrm{dip}}) \chi_2\\
\label{delta} \widetilde{\delta}&=& \delta_{\textrm{dip}}\chi_1 + (\delta'_{\textrm{tr}}+\delta'_{\textrm{dip}}) \chi_2\\
\label{theta} \widetilde{\vartheta}&=& \vartheta_{\textrm{dip}} \chi_1+(\vartheta'_{\textrm{tr}}+\vartheta'_{\textrm{dip}}) \chi_2~,
\eea
where $\chi_1=\text{M}_1/\text{H}_0$ and $\chi_2=\text{M}_2/\text{H}_0$ denote the local susceptibilities per Cu$_1$ and Cu$_2$ respectively.

\subsection{Comparison with experiment}\label{comparison}
Let us now compare the theoretical predictions given in Table \ref{TabSeCoordBloc} to our experimental data.
We find exact agreement for the number of distinct lines as well as for their relative intensities.
When the magnetic field is applied along [111] our model predicts that the group of Se$_{\text{I}}$ sites give
one spectral line for Se$_\textrm{I}^1$ and a separate, three times more intense line from Se$_\textrm{I}^2$, Se$_\textrm{I}^3$ and Se$_\textrm{I}^4$.
A similar result holds for the group of Se nuclear spins of the second type.
In the [110] direction we expect four resonance lines with relative ratio 2:2:2:2, while in the [100] direction we expect two lines with relative ratio 4:4 (see Table~\ref{TabSeCoordBloc}).
These predictions are in perfect agreement with our experimental results shown in Figs.~\ref{Lines220K} and \ref{LinesVsT}.

\begin{figure}[!t]
\centering
\includegraphics[width=0.48\textwidth]{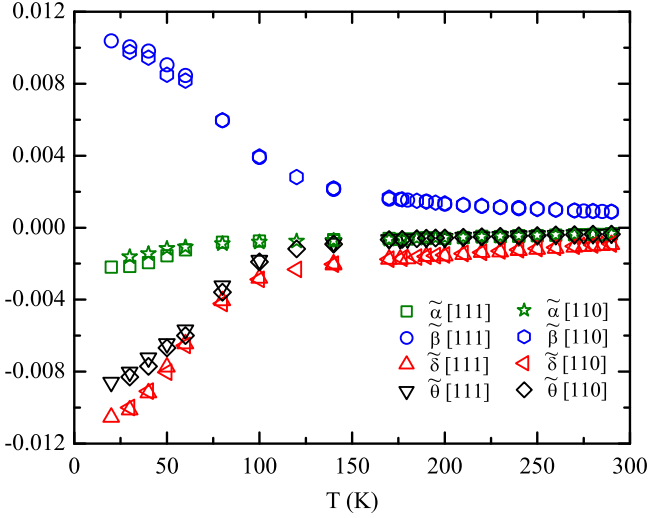}
\caption{(Color online) A comparison of the values of the parameters given in Eqs. (\ref{alpha})-(\ref{theta})
extracted from the NMR line shifts along the [111] direction with those extracted from the [110] direction.}
\label{tilda}
\end{figure}

For consistency reasons we would like next to contrast the T-dependence of $\widetilde{\alpha}$, $\widetilde{\beta}$, $\widetilde{\delta}$, and $\widetilde{\vartheta}$
obtained from the NMR data along the [111] direction with that obtained from the NMR data along the [110] direction.
The results (taken after correcting the data for the chemical shift and the demagnetization field, cf. below) are shown in Fig.~\ref{tilda} and are almost identical.
This provides a much stronger confirmation of the internal consistency of the above theory.
In addition, Fig.~\ref{tilda} tells us that we may use either the [111] or the [110] data in order to extract, in conjunction with Eqs.~(\ref{alpha})-(\ref{theta}),
the transferred hyperfine field parameters as well as the T-dependence of the local moments M$_1$ and M$_2$. In what follows we have taken the data along [111].
\begin{table}[!t]
\caption{Extracted estimates of the components of the transferred hyperfine tensors $\vec A^{\textrm{Cu}_1}_{\textrm{Se}^n_{\textrm{I}}}$, $\vec A^{\textrm{Cu}_{2}}_{\textrm{Se}^n_\textrm{I}}$ and  $\vec A^{\textrm Cu_2}_{\textrm Se^n_\textrm{II}}$ defined in Eqs. (\ref{Se1Cu1Cu2matrices}) and (\ref{Se2Cu2matrice}) (here in units of $10^{22} \textrm{cm}^{-3}$) for two possible values
of the constant $r_1$ defined in the text.}\label{valuesHIT}
\begin{ruledtabular}
\begin{tabular}{c c c c c c c}
$r_{1}$   & $\alpha_{\textrm{tr}}$ & $\beta_{\textrm{tr}}$& $\alpha'_{\textrm{tr}}$& $\beta'_{\textrm{tr}}$ & $ \delta'_{\textrm{tr}}$ & $\theta'_{\textrm{tr}}$ \\
\hline
$-2$ &$-2.3$ & $12.6$ & $-5.0 $& $26.8 $ & $-14.6$ & $-14.0$  \\
$-1.2$&$-1.7$ &$-5.4$ &$-1.2$ & $-3.8$ &$-16.8$ & $-16.2$  \\
\end{tabular}
\end{ruledtabular}
\end{table}

\textit{Assignment of the NMR lines}.---
The next step is to identify the specific subgroup of Se sites associated to each given NMR line.
To this end, we first note that the local fields of Se$^n_\textrm{I}$ and Se$^n_\textrm{II}$
given in Table~\ref{TabSeCoordBloc} map to one another when interchanging $(\widetilde{\alpha}, \widetilde{\beta}) \mapsto (\widetilde{\delta}, \widetilde{\vartheta})$.
This prevents a straightforward assignment of the NMR lines to specific subgroup of selenium sites.
For example, we cannot infer if it is P1 and P2 or rather P1 and P4 (cf. Fig.~\ref{LinesVsT}) that come from the same type of Se sites.
We overcome this drawback by comparing the fits in different directions of the applied field.
Indeed, by fitting the experimental data of $\Delta\nu^{exp}_{[111]}$ (see Table~\ref{Table220K}) with the theoretical expressions given in Table~\ref{TabSeCoordBloc}
we can reproduce the experimental data of both $\Delta\nu^{exp}_{[110]}$ and $\Delta\nu^{exp}_{[100]}$ only under the condition
that P1 and P2 come from one group of selenium sites while P3 and P4 come from the other.
This result is further supported by the nuclear spin-lattice and spin-spin relaxation time measurements presented in the following section.
However it is still not possible at this point to tell whether the pair P1-P2 comes from type-I or type-II Se sites.
We have extracted the temperature dependence of the local moments using both possibilities and we have found that the choice which
gives the most physically reasonable behavior is the one which assigns the pair P1-P2 to type-II Se sites.

\begin{figure}[!t]
\centering
\includegraphics[width=0.45\textwidth]{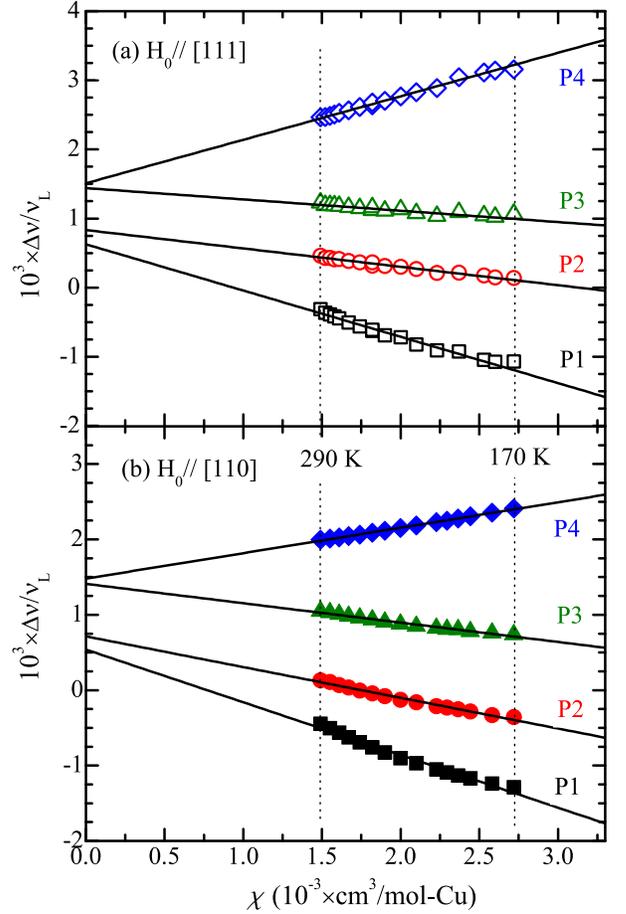}
\caption{(Color online) The fractional shift $\Delta\nu_{\textrm{P}_{i}}/\nu_{\textrm{L}}$ for each one of the four NMR peaks as measured along (a) [111] and (b) [110], vs. the susceptibility $\chi$ measured with the magnetometer with $T$ being an implicit parameter. The dotted lines indicates the points where $T= 170$ K, and $290$ K.}
\label{fraction}
\end{figure}

\textit{Hyperfine parameters and local susceptibilities}.---
We are now ready to extract the local moments from the NMR data. We first write the measured susceptibility per Cu and the measured relative shifts $K\equiv\Delta\nu /\nu_L$ as
\bea
\label{chiEq}
\chi_\textrm{tot}(T)&=&\chi_{\text{spin}}(T) + \chi_{\text{dia}}+\chi_{\text{vv}}\\
\label{KEq}
K(T)&=&K_\text{spin}(T)+K^c~.
\eea
Here $\chi_\textrm{spin}=(\chi_1+3\chi_2)/4$ is the total spin susceptibility per Cu, and $\chi_\text{dia}$, $\chi_\text{vv}$ are the T-independent diamagnetic and van Vleck contributions respectively. Similarly, $K_\text{spin}(T)$ includes the dipolar and the transferred hyperfine contributions which were given above in Table \ref{TabSeCoordBloc}, while $K^c$ stands for the T-independent chemical shift.

\begin{figure}[!t]
\centering
\includegraphics[width=0.49\textwidth]{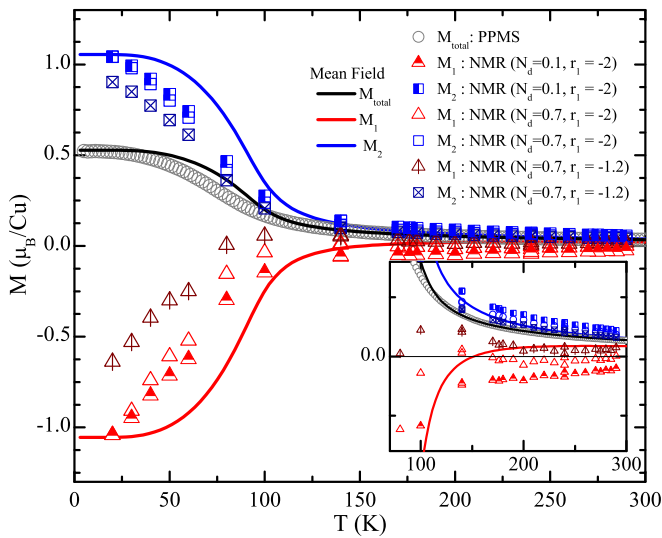}
\caption{(Color online) Comparison between the total magnetization measured by PPMS (corrected for the van Vleck and diamagnetic contributions) and the local moments extracted from NMR (with two different values of the ratio $r_1$, cf. text) with the corresponding mean field theory predictions. We also demonstrate how the value of the demagnetizing factor $N_d$ affects the sign of M$_1$ at high temperatures.}
\label{MFvsNMR}
\end{figure}

Figure \ref{fraction}(a) shows the so-called Clogston-Jaccarino plot\cite{CJ} of the four measured shifts $K$ in both [111] and [110] directions versus the measured susceptibility per mol Cu in the high temperature regime. The linear behavior shows that both $\chi_1$ and $\chi_2$ are proportional to $\chi_\text{spin}$ in this regime. We can exploit this linear behavior in order to extract the values of $K^c$ for each line if we make an estimate of $\chi_\text{dia}+\chi_\text{vv}$. For the latter we take the upper bound of $\chi_\text{dia}+\chi_\text{vv} \simeq 1\times10^{-4}$ cm$^3$/mol-Cu which is obtained from the comparison of the measured magnetization with the mean field theory prediction (cf. below).
We extract the following estimates respectively for the lines P1, P2, P3 and P4: $K^c\simeq 5.60 \times 10^{-4}$, $8.06\times 10^{-4}$, $14.2\times 10^{-4}$, $15.7\times 10^{-4}$ along [111],
while $K^c\simeq 4.68\times 10^{-4}$, $6.73\times 10^{-4}$, $13.8\times 10^{-4}$, $15.1\times 10^{-4}$  along [110].

After subtracting the above values of $\chi_\text{dia}+\chi_\text{vv}$ and $K_\text{chem}$ from the bare data, we adopt the following procedure.
We first extract an estimate for $\theta_\text{tr}'$ by using the data for $\widetilde{\theta}$ at the lowest available temperature ($T=20$ K) and the relation M$_1(20 K) = r_1\text{M}(20 K)$, where $r_1$ is a constant which measures the effect of quantum fluctuations at low temperatures. Using Eq. (\ref{theta}), the relation M$=$(M$_1+3$M$_2)/4$, and the PPMS data for M, we may extract the whole temperature dependence of M$_1$ and M$_2$. In turn, Eqs. (\ref{alpha})-(\ref{delta}) can provide the remaining transferred hyperfine parameters in a straightforward way.
For comparison, we have used two values of $r_1$: The first, $r_1=-2$, corresponds to the case without quantum fluctuation effects, while the second value, $r_1=-1.2$, corresponds to the reduction in the spin length found by Neutron diffraction data.\cite{Bos}
As for the demagnetization factor, it turns out that for $N_d \lesssim 0.6$ the extracted M$_1(T)$ is not positive at room temperatures but saturates to a negative value.
However such a behavior would be quite unphysical:
Although there is a large antiferromagnetic exchange field on each Cu$_1$ site since it has 6 neighboring Cu$_2$ sites, one still expects that M$_1$ should ultimately change sign at some temperature $T^*$ which is above $T_c$ but certainly well below room temperature. On this issue, the mean-field theory described below predicts that $T^*\simeq 152$ K. So we adjust $N_d \simeq 0.7$, which in conjunction with the value of $r_1$ can provide a physically more reasonable behavior at high temperatures.

The resulting estimates of the hyperfine parameters are provided in Table \ref{valuesHIT}, while the extracted T-dependence of the local moments is shown in Fig.~\ref{MFvsNMR}.
The major result contained in Fig.~\ref{MFvsNMR} is that the NMR data are fully consistent with the 3up-1down ferrimagnetic state, namely that the Cu$_2$ moments are aligned parallel and the Cu$_1$ moments are aligned antiparallel to the applied field. Hence NMR on single crystals provides a more direct local probe of this state, mapping out the whole T-dependence of the local moments.

{\it Comparison to microscopic model}:---
It is worthwhile to contrast the above NMR findings with a microscopic spin model for Cu$_2$OSeO$_3$. Following the Kanamori-Goodenough rules (see also discussion in Ref. \onlinecite{Bos}) we introduce two exchange parameters: one ferromagnetic ($J_\text{fm}$) between Cu$_2$-Cu$_2$ ions and one antiferromagnetic ($J_\text{afm}$) between Cu$_1$-Cu$_2$ ions.
Looking at the structure one finds that each Cu$_1$ neighbors six Cu$_2$ ions while each Cu$_2$ neighbors four Cu$_2$ ions and two Cu$_1$ ions.
The self-consistent equations of the corresponding mean field theory are given by
\bea
\label{mfm1}
m_1 &=& \text{tanh}\left( \frac{b -3 J_\text{afm} m_2}{2T} \right)\\
\label{mfm2}
m_2 &=& \text{tanh}\left( \frac{b +2|J_\text{fm}| m_1-J_\text{afm} m_2}{2T} \right)
\eea
where $m_{1,2}\equiv\text{M}_{1,2}/(g\mu_B/2)$, $b\equiv g\mu_B H_0/k_B$, and the exchange constants are in units of Kelvin.
The mean field transition temperature is
$T_c^\text{MF}=\frac{1}{2}\left( |J_\text{fm}|+\sqrt{J_\text{fm}^2+3J_\text{afm}^2} \right)$,
while the predictions for the local and the total spin susceptibility per Cu at $T>T_c$ are given by the expressions
\bea
\chi_1&=&\frac{(g\mu_B)^2}{8k_B} \frac{ 2T-2|J_\text{fm}|-3J_\text{afm} }{ (T-T_c)(T+T_c-|J_\text{fm}|) }\\
\chi_2&=&\frac{(g\mu_B)^2}{8k_B} \frac{ 2T-J_\text{afm} }{ (T-T_c)(T+T_c-|J_\text{fm}|) }\\
\chi_\text{spin}&=&\frac{(g\mu_B)^2}{8k_B} \frac{ 2T-1/2|J_\text{fm}|-3/2J_\text{afm} }{ (T-T_c)(T+T_c-|J_\text{fm}|) } .
\eea
Using the exact numerical solution of Eqs. (\ref{mfm1}) and (\ref{mfm2}) we have obtained a quite accurate fit (cf. Fig. \ref{MFvsNMR}, solid black line) of the measured magnetization data in the range 110-300 K, using $g\simeq 2.11$,\cite{Larranaga} $J_\text{afm}\simeq 68$ K and $J_\text{fm}\simeq -50$ K. This fit puts an upper bound on the diamagnetic and van Vleck contributions to the susceptibility $\chi_\text{dia}+\chi_\text{vv} \simeq 1\times10^{-4}$ cm$^3$/mol-Cu which is of the right order of magnitude.
Figure \ref{MFvsNMR} shows also the solution for the temperature dependence of the local magnetizations M$_1$ and M$_2$ (solid red and blue lines) which are to be contrasted with the
behavior extracted from NMR. The agreement is quite satisfactory.
Of particular interest is the behavior of M$_1$ at high temperatures. As we discussed above, the Cu$_1$ moments remain antiparallel to the field even above $T_c$ due to the large negative exchange field that is exerted from the 6 neighboring Cu$_2$ moments. The mean field theory prediction for the temperature $T^{*}$ at which M$_1$ eventually changes sign is given by $T^{*} = \frac{3}{2} J_\text{afm} + |J_\text{fm}| \simeq 152$ K.

So the mean field theory provides a semi-quantitative agreement with the measured magnetization data and captures the essential local physics of the problem, being in agreement with the
picture obtained from NMR. On the other hand, the mean-field theory does not provide a good description at temperatures close to $T_c$ since it neglects long-range correlation effects.
In addition it can overestimate the value of $T_c$, and does not capture quantum fluctuation effects which give rise to an overall reduction in the length of the local moments at low enough temperatures.

\section{nuclear spin-lattice and spin-spin relaxation rates}\label{dynamic}
The $^{77}$Se spin-lattice and spin-spin relaxation times,  $T_1$ and $T_2$ respectively, were measured for the four different peaks of the spectrum
when the magnetic field is parallel to the [111] axis.
We first note that the lines P1 and P2 have very similar $T_1$ and $T_2$ values over the whole range of T's investigated here and the same holds true for the lines P3 and P4. This is yet another confirmation that the lines P1-P2 belong to one type of Se sites and the lines P3-P4 belong to the second type.

\begin{figure}[!t]
\centering
\includegraphics[width=0.47\textwidth]{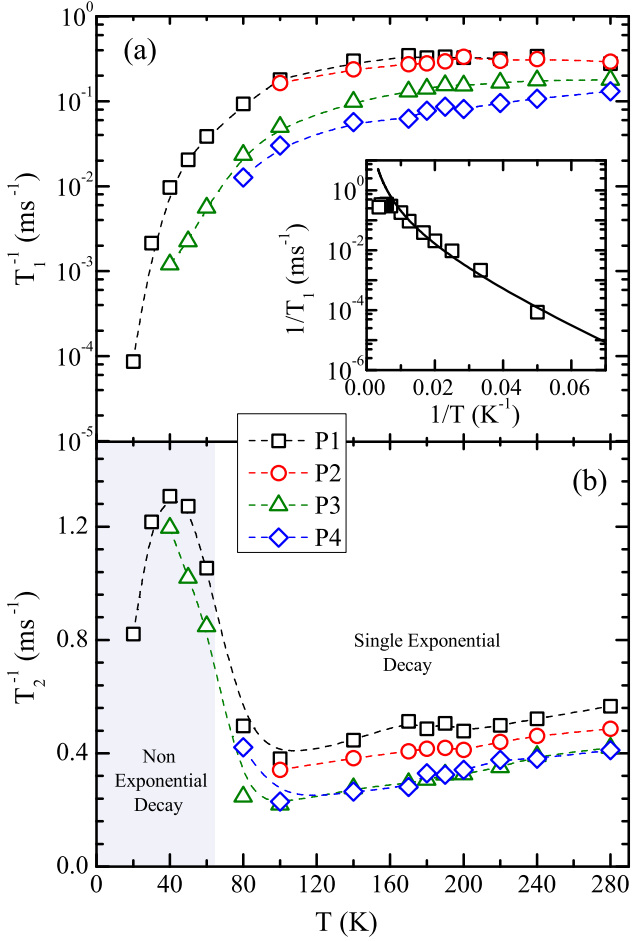}
\caption{(Color online) Temperature dependence of the $^{77}$Se NMR (a) spin-lattice $T_1^{-1}$ and (b) spin-spin $T_2^{-1}$ relaxation rates
with the magnetic field of 14.09 T applied parallel to [111]. The inset shows $1/T_1$ vs. 1/T for the line P1 and the corresponding fit (solid line) with the contribution from Raman scattering of magnons.}
\label{T1T2}
\end{figure}

The recovery of the longitudinal magnetization follows, for each peak, a single exponential behavior in time, as expected for a $I=1/2$ nuclear spin. The values of spin-lattice relaxation rate $1/T_1$ derived from the fit of the recovery data are shown in Fig.~\ref{T1T2}(a).
At $T>$ 170 K, $1/T_1$ varies weakly with temperature for all lines.
We may obtain a theoretical fit of $1/T_1$ by considering the Raman scattering of magnons off the $^{77}$Se nuclei
which is expected to be the dominant process (at low enough temperatures) that conserves both the total energy and the total angular momentum of the nuclear+electron spin system. To this end, we shall make the reasonable assumption that the relevant lowest spin-wave excitation branch around the minimum
$\Delta$ (which is set by the external field)
has a typical parabolic dispersion of the form $\epsilon_\vec{k}= \Delta + \lambda (a\vec{k})^2$, where $a$ is the lattice constant
and $\lambda$ is an energy scale of the order of the actual exchange couplings in the system which gives the curvature around the minimum of the lowest magnon band.
Following similar arguments with Refs. [\onlinecite{Moriya56,Beeman68}] one obtains
\be\label{T1Eq}
\frac{1}{T_1} = z \frac{(A/z)^2 \text{sin}^2\theta}{2^5 \pi^3 \hbar \lambda} \left(\frac{k_BT}{\lambda}\right)^2\int_{\Delta/k_BT}^\infty \frac{dx}{e^x-1}
\ee
where $z$ is the number of Cu$^{2+}$ ions closest to the nuclear spin, $A/z$ stands for the hyperfine coupling with a single electronic spin,
and $\theta$ is the angle between the quantization axes of the electronic and the nuclear spin.
Our optimal fit based on a numerical evaluation of the integral of Eq.~(\ref{T1Eq}) with $\Delta/k_B= g\mu_B H$ ($H=14$ T)
is shown in the inset of Fig.~\ref{T1T2}(a) (solid line) and gives $A^2\text{sin}^2\theta/\lambda^3 \simeq 1.6\times 10^7$ erg$^{-1}$.
Taking $A/q \sim 2\times 10^{23}$ cm$^{-3}$ for the line P1, gives $\lambda/k_B\sim 50$ K which is of the right order of magnitude.

Figure~\ref{T1T2}(b) shows the T-dependence of the spin-spin relaxation rate $1/T_2$.
These results were obtained by fitting the decay of the spin echo signal $M(2\tau)$, after a $\pi/2-\tau-\pi$ pulse sequence, with the proper functional form.
We have found that the decay of the transverse nuclear magnetization follows a single exponential decay in the range 80-290 K
while it deviates from the single exponential law below 60 K. The irreversible decay $M(2\tau)$ of the $^{77}$Se echo signal has the following contributions.
First, the $^{77}$Se nuclear dipole-dipole interaction which can be estimated from the calculation of the van Vleck second moment\cite{Vleck}
$M_2^{\text{Se-Se}}$ taking into account the natural abundance of $^{77}$Se.\cite{Abragam} We have found that $\sqrt{M_2^{\text{Se-Se}}}\approx85 s^{-1}$ with a slight variation among the different Se sites, which is much smaller than the experimental values reported in Fig.~\ref{T1T2}.
The second contribution to the spin echo decay is the Redfield term,\cite{Slichter96} which is of the form $\text{exp}(-2\tau/T_1)$. From the $1/T_1$ values reported in Fig.~\ref{T1T2}(a) it is obvious that the contribution of the Redfield term in the spin echo decay is also very small here.

The third contribution to the spin echo decay is due to fluctuating dipolar fields between unlike nuclear spins with the most dominant being the one among
Se and Cu nuclei.
Depending on the time scale (or correlation time $\tau_c$) of these fluctuating dipolar fields we have the following two limiting regimes:
(i) The fast motion regime, characterized by $\gamma_N\langle\delta H_z\rangle\tau_c\ll 1$, where the spin echo decay is a single exponential, $\text{exp}(-2\tau/T_2)$, with $1/T_2=\gamma_N^2\langle\delta H_z^2\rangle\tau_c$,\cite{Slichter96} and $\gamma_N^2\langle\delta H_z^2\rangle$ is the second moment of the selenium-copper dipolar interaction for which a straightforward lattice sum gives $M_2^{\text{Se-Cu}}\simeq 7.7\times10^6 s^{-2}$.
The fast motion approximation is in our case applicable for $T\gtrsim 80 K$.
(ii) The quasi-static regime, characterized by $\gamma_N\langle\delta H_z\rangle\tau_c \gg 1$, where the slow fluctuations of the selenium-copper dipolar interaction give rise to a non-exponential spin echo decay. In the nearly static regime the decay goes as $\text{exp}[-(2\tau/T_2)^3]$, where $1/T_2=(\gamma_N^2\langle\delta H_z^2\rangle/12\tau_c)^{1/3}$.\cite{Takigawa} Identifying $\tau_c$ with the spin-lattice relaxation time of Cu nuclei, which at 30 K is measured to be about 1ms,\cite{T1Cu} we find $1/T_2\simeq 862~s^{-1}$ in agreement with our low-T data. As we noted above the decay of the transverse magnetization is not exponential at low T's. A good fit of the data can be obtained by both a square exponential decay (half Gaussian) and by a third power exponential. The discrimination between them is prevented by the weakness of the NMR signal at long times.

\section{Summary}\label{concl}
We have presented extensive $^{77}$Se NMR measurements in single crystals of the magnetoelectric ferrimagnet Cu$_2$OSeO$_3$.
The analysis of the data has provided a number of central findings both for the crystalline as well as the magnetic phase of this compound.
First, the T-dependence of the two types of Cu$^{2+}$ moments, extracted from the NMR data, is fully consistent with a phase transition
from the high-T paramagnetic phase to a low-T ferrimagnet whereby 3/4 of the Cu$^{2+}$ moments are aligned parallel and 1/4 antiparallel to the applied field.
Below the transition temperature we do not observe any clear change in the broadening of the NMR lines or any splitting of the NMR lines, which shows that there is no measurable symmetry reduction in the crystalline structure from its high-T space group \textit{P}2$_1$3. These results are in agreement with previous data from magnetization, high-resolution x-ray, and Neutron diffraction measurements on powder samples reported by Bos {\it et al,}\cite{Bos} but also from infrared,\cite{Miller} and Raman\cite{Gnezdilov} studies in single crystals of Cu$_2$OSeO$_3$.

We have also developed a microscopic spin model with two nearest-neighbor exchange interactions: One antiferromagnetic ($J_\text{afm}\simeq 68$ K) between Cu$_1$ and Cu$_2$ ions which forces them to be antiparallel to each other, and a second ferromagnetic interaction between nearest-neighbor Cu$_2$ ions which aligns then in the same direction, giving rise to the above
ferrimagnetic state. We have shown that a mean field solution of this model provides a very good description of the physics of the problem and is in excellent agreement with measured magnetization data in a wide temperature range and, more importantly, it is also consistent with the local picture extracted from NMR. A first-principles study of this system may provide a more accurate and refined microscopic spin model.\cite{oleg}

More generally, our NMR study provides a strong local confirmation that the ferrimagnetic ordering in Cu$_2$OSeO$_3$ does not proceed via a spontaneous lattice distortion and thus
this material provides a unique example of a metrically cubic crystal that allows for piezoelectric as well as linear magnetoelectric and piezomagnetic coupling.

\section{Acknowledgments}
We would like to thank K. Schenk, O. Janson, A. Tsirlin, L. Hozoi, and M. Abid for useful discussions.
We also acknowledge experimental assistance from K. Schenk, M. Zayed, P. Thomas, and S. Granville.
One of the authors (H. B.) acknowledges financial support from the Swiss NSF and by the NCCR MaNEP.


\begin{thebibliography}{99}
\bibitem{Eerenstein} W. Eerenstein, N. D. Mathur, and J. F. Scott, Nature \textbf{442}, 759 (2006).
\bibitem{Spaldin} N. A. Spaldin, M. Fiebig, Science \textbf{309}, 391 (2005).
\bibitem{Khomskii} D. Khomskii, Physics \textbf{2}, 20 (2009).
\bibitem{Wang} K. F. Wang, J.-M. Liu, and Z. F. Ren, Advances in Physics \textbf{58}, 321 (2009).
\bibitem{Landau} L. D. Landau, and E. M. Lifshitz, Electrodynamics of continuous media (Pergamon Press, Oxford, 1984). 
\bibitem{Dzyaloshinskii} I. E. Dzyaloshinskii, Sov. Phys. JETP \textbf{10}, 628 (1959).
\bibitem{Gajek} M. Gajek, M. Bibes, S. Fusil, K. Bouzehouane, J. Fontcuberta, A. Barth\'el\'emy, and A. Fert, Nature Materials \textbf{6}, 296 (2007).
\bibitem{Wood} V. E. Wood, and A. E. Austin, Int. J. Magn. \textbf{5}, 303 (1974).
\bibitem{Fiebig} M. Fiebig, J. Phys. D: Appl. Phys. \textbf{38}, R123 (2005).
\bibitem{Schmid} H. Schmid, Ferroelectrics \textbf{162}, 317 (1994).
\bibitem{Folen} V. J. Folen, G. T. Rado, and E. W. Stalder, Phys. Rev. Lett. \textbf{6}, 607 (1961).
\bibitem{Wiegelmann} H. Wiegelmann, A. A. Stepanov, I. M. Vitebsky, A. G. M. Jansen, and P. Wyder, Phys. Rev. B \textbf{49}, 10 039 (1994).
\bibitem{Fox} D. L. Fox, J. F. Scott, J. Phys. C  \textbf{10}, 329 (1977).
\bibitem{Kimura} T. Kimura, T. Goto, H. Shintani, K. Ishizaka, T. Arima, and Y. Tokura, Nature \textbf{426}, 55 (2003).
\bibitem{Hur} N. Hur, S. Park, P. A. Sharma, J. S. Ahn, S. Guha, and S.-W. Cheong, Nature \textbf{429}, 392 (2004).
\bibitem{Lawes0} G. Lawes, A. B. Harris, T. Kimura, N. Rogado, R. J. Cava, A. Aharony, O. Entin-Wohlman, T. Yildrim, M. Kenzelmann, C. Broholm, and A. P. Ramirez, Phys. Rev. Lett. \textbf{95}, 087205 (2005).
\bibitem{Lautenschl} G. Lautenschl\"{a}ger, H. Weitzel, T. Vogt, R. Hock, A. B\"{o}hm, M. Bonnet, and H. Fuess, Phys. Rev. B \textbf{48}, 6087 (1993).
\bibitem{Taniguchi} K. Taniguchi, N. Abe, T. Takenobu, Y. Iwasa, and T. Arima, Phys. Rev. Lett. \textbf{97}, 097203 (2006).
\bibitem{Choi} Y. J. Choi, H. T. Yi, S. Lee, Q. Huang, V. Kiryukhin, and S.-W. Cheong, Phys. Rev. Lett. \textbf{100}, 047601 (2008).
\bibitem{Pregelj} M. Pregelj, O. Zaharko, A. Zorko, Z. Kutnjak, P. Jegli\v{c}, P. J. Brown, M. Jagodi\v{c}, Z. Jagli\v{c}i\'{c}, H. Berger, and D. Ar\v{c}on, Phys. Rev. Lett. \textbf{103}, 147202 (2009).
\bibitem{Lawes} G. Lawes, A. P. Ramirez, C. M. Varma, and M. A. Subramanian, Phys. Rev. Lett. \textbf{91}, 257208 (2003).
\bibitem{Bos} J-W. G. Bos, C. V. Colin, and T. T. M. Palstra, Phys. Rev. B \textbf{78}, 094416, (2008).
\bibitem{Miller} K. H. Miller, X. S. Xu, H. Berger, E. S. Knowles, D. J. Arenas, M. W. Meisel, and D. B. Tanner, arXiv:1006.467v1 [cond-mat.mtrl-sci].
\bibitem{Gnezdilov} V. P. Gnezdilov, K. V. Lamonova, Yu. G. Pashkevich, P. Lemmens, H. Berger, F. Bussy, and S. L Gnatchenko, Fizika Nizkikh Temperatur \textbf{36}, 688 (2010). 
\bibitem{Effenberger} H. Effenberger, and F. Pertlik, Monatsh. Chem. \textbf{117}, 887, (1986).
\bibitem{Larranaga} A. Larra\~{n}aga, J. L. Mesa, L. Lezama, J. L. Pizarro, M. I. Arriortua, and T. Rojo, Materials Research Bulletin, \textbf{44}, 1, (2009).
\bibitem{IntTablesCryst} International Tables for Crystallography (2006). Vol. A, ch. 7.1, pp. 610–611 (2006).
\bibitem{White} R. M. White, Quantum Theory of Magnetism (McGraw-Hill, New York, 1970).
\bibitem{CJ} A. M. Clogston, and V. Jaccarino, Phys. Rev.  {\bf 121}, 1357 (1960).
\bibitem{Moriya56} T. Moriya, Prog. Theor. Phys. \textbf{16}, 23 (1956); \textbf{16}, 641 (1956).
\bibitem{Beeman68}  D. Beeman and P. Pincus, Phys. Rev. {\bf 166}, 359 (1968).
\bibitem{Vleck} J. V. Vleck, Phys. Rev. \textbf{74}, 1168 (1948).
\bibitem{Abragam} A. Abragam, Principles of Nuclear Magnetism (Oxford University press, 1983).
\bibitem{Slichter96} C. P. Slichter, Principles of Magnetic Resonance (Springer-Verlag, 1996).
\bibitem{Takigawa} M. Takigawa, G. Saito, J. Phys. Soc. Jpn. \textbf{55}, 1233 (1986).
\bibitem{T1Cu} The copper NMR signal (not shown here) could be observed in Cu$_2$OSeO$_3$ only below 30 K.
\bibitem{oleg} O. Janson, and A. Tsirlin (private communication).
\end{thebibliography}
\end{document}